\documentclass[useAMS, usenatbib, usegraphicx]{mn2e}
\usepackage{amssymb}
\usepackage{booktabs}
\newcommand\AnA{A\&A}

\def\alf{Alfv\'en\,}
\def\bq{\begin{equation}}
\def\eq{\end{equation}}
\def\ee #1 {\times 10^{#1}}
\def\ut #1 #2 { \, \rmn{#1}^{#2}}
\def\u #1 { \, \rmn{#1}}

\let\grad=\nabla
\newcommand\cross{\bmath{\times}}

\newcommand\bp{b_\perp}
\def\curl{{\grad \cross}}
\def\div #1 {\grad \cdot #1}

\def\cs{\cos\theta}

\def\b{\bmath{b}}

\def\v{\bmath{v}}
\def\duz{\delta u_z}
\def\dux{\delta u_x}
\def\duy{\delta u_y}
\def\dup{\delta u_+}
\def\dum{\delta u_-}

\def\k{\bmath{k}}
\def\kh{\hat{\k}}

\def\x{\bmath{x}}
\def\vi{\bmath{v}_i}

\def\ve{\bmath{v}_e}
\def\vi{\bmath{v}_i}

\def\vn{\bmath{v}_n}

\def\J{\bmath{J}}
\def\B{\bmath{B}}
\def\dBx{\delta B_x}
\def\dBy{\delta B_y}
\def\dBz{\delta B_z}
\def\dBpl{\delta B_+}
\def\dBm{\delta B_-}
\def\E{\bmath{E}}            % E
\def\hB{\hat{\B}}
\def\Bz{B_z}
\def\vz{v_z}

\def\J{\bmath{J}}

\def\drho{\delta\rho}
\def\dv{\bmath{\delta\v}}

\def\dB{\bmath{\delta\B}}

\def\eh{\hat{\eta}_H}
\newcommand{\delt} [1] {\frac{\partial #1}{\partial t}}

\newcommand{\delz} [1] {\frac{\partial #1}{\partial z}}

\def\b1{{\bar{\omega}}}
\def\bo2{\bar{\omega}^2}
\newcommand{\dxi} [1] {\frac{\partial #1}{\partial \xi}}

\newcommand{\dtau} [1] {\frac{\partial #1}{\partial \tau}}
\newcommand{\dt} [1] {\frac{\partial #1}{\partial t}}
\newcommand{\dpz} [1] {\frac{\partial #1}{\partial z}}

\begin{document}

\title{Waves in the solar photosphere}
\author[B. P. Pandey, J. Vranjes and V. Krishan]{B. P. Pandey$^{1}$
\thanks{E-mail:bpandey@physics.mq.edu.au; jovo.vranjes@wis.kuleuven.be}, and J. Vranjes$^2$ and V. Krishan$^3$
\\
$^{1}$Department of Physics, Macquarie University, Sydney 2109, Australia\\
$^{2}$Centre for Plasma Astrophysics, Celestijnenlaan 200B, 3001 Leuven, Belgium,\\ and Facult\'{e} des Sciences Appliqu\'{e}es, avenue F.D. Roosevelt 50,  1050 Bruxelles, Belgium\\
$^3$Indian Institute of Astrophysics, Bangalore 560034, India}
\date{\today}
\pagerange{\pageref{firstpage}--\pageref{lastpage}}
\pubyear{2007}
\maketitle
\label{firstpage}
\begin{abstract}
\maketitle 
The solar photosphere is a partially ionized medium with collisions between electrons, various metallic ions and neutral hydrogen playing an important role in the momentum and energy transport in the medium. Furthermore, the number of neutral hydrogen atom could be as large as $10^{4}$ times the number of plasma particles in the lower photosphere. The non-ideal MHD effects, namely Ohm, Ambipolar and Hall diffusion can play an important role in the photosphere. We demonstrate that Hall is an important non-ideal MHD effect in the solar photosphere and show that Hall effect can significantly affect the excitation and propagation of the waves in the medium.  We also demonstrate that the non-ideal Hall dominated inhomogeneous medium can become parametrically unstable, and it could have important ramification for the photosphere and chromosphere of the sun. The analysis hints at the possibility of solar photosphere becoming parametrically unstable against the linear fluctuations. 
\end{abstract}


\section{Introduction}
The solar atmosphere is a partially ionized medium with fractional ionization 
($X_e = n_e/n_n$) -- a ratio of the electron ($n_e$) to the neutral ($n_n$) number densities, changing 
with the altitude. In fact, the 
lower solar atmosphere -- the photosphere ($\le 500\,\mbox{km}$) is a weakly ionized medium where collisions 
between electrons, metallic ions and neutrals are responsible for the dynamical behaviour of the medium. With 
increasing altitude, the fractional ionization, which is $10^{-4}$ in the photosphere drops to $10^{-2}$ in the chromosphere at 
$1000  \,\mbox{km}$. Even at an altitude $\sim \left(2000 - 2500 \right)\,\mbox{km}$, the neutral number density 
dominates the plasma density by an order 
of magnitude \citep{VAL81}. Thus, the solar atmosphere is a partially ionised mixture of plasma and neutral gas 
with very weakly ionized gas in the photosphere and fully ionized plasma in the upper chromosphere and beyond. 

Clearly, a partially ionized mixture of gas does not behave like fully ionized plasma and hence, ideal magnetohydrodynamics (MHD) description of the dynamics of solar photosphere at best is a very crude approximation. 
The investigation of \alf wave in the ideal MHD has been very popular topic in the solar physics owing to its appeal to 
the solar coronal heating \citep{I78, P87, GP04}. These waves can be easily excited in the medium due to easy availability of 
vast reservoir of energy such as the convective gas motions. Only a tiny fraction of this energy, if carried by the 
wave to higher altitude, will suffice to heat the corona to high temperatures \citep{PPG97}. Implicit in this picture is the 
assumption that non-ideal MHD effects are unimportant in the photosphere. However, collision effects may severly affect 
the propagation of the wave in the solar atmosphere \citep{DPH98, G98, G04, LAK05, VPP07, AHL07, VPPD07}. 

The importance of non-ideal MHD effect is related to the question of how well the magnetic field is coupled to the neutral 
matter. These effects can be quantified in terms of plasma--Hall parameter $\beta_j$
\bq
\beta_j = \frac{\omega_{cj}}{\nu_{j}}\,,
\eq
a ratio of the $\mbox{j}^{\mbox{th}}$ particle cyclotron frequency $\omega_{cj} = e\,B/m_j\,c$ (where
$e\,,B\,,m_j\,,c$ denots the electron charge, magnetic field, mass and speed of the light respectively) to the 
sum of the plasma-plasma, and plasma -- neutral,  $\nu_{jn}$collision frequencies. For 
electrons $\nu_{e} = \nu_{en} + \nu_{ei} + \nu_{ee} \approx 2\,\nu_{en} + \nu_{ei}$ and for ions 
$\nu_{i} = \nu_{in} + \nu_{ii}$. It is clear that whenever $\beta_j \gg 1$, the $\mbox{j}^{\mbox{th}}$ plasma particle will be 
strongly tied to the ambient magnetic field of the medium whereas in the opposite case, particles will not feel the 
Lorentz force and thus, can be treated as weakly magnetized or unmagnetized. For example, when ions and neutrals are strongly 
coupled, the relative drift of 
the ``frozen-in" ions ($\beta_i \gg 1$) against the sea of neutrals may cause the diffusion of the magnetic 
flux -- the so called ambipolar diffusion. When the ions and neutrals are moving together ($\beta_i \lesssim 1$), and there is 
relative drift between the electrons and ions (i.e. electrons are ``frozen-in" the field but ions are not), the 
Hall diffusion
\footnote{We note that the Hall drift could be a more appropriate terminology as diffusion is often associated with the dissipation which for Hall diffusion is zero. However, we shall use Hall diffusion to maintain the uniformity of the description.} becomes important. When even electrons are not ''frozen--in'' the field, i.e. $\beta_e \lesssim 1$, the 
Ohmic dissipation becomes important. Therefore, the investigation of the non-ideal magnetohydrodynamic (MHD) effect 
in the solar atmosphere should be investigated in different parameter windows:  (I) ambipolar -- when the magnetic field 
can be regarded as frozen in 
the plasma and drifts with it through the neutrals (II) Ohmic -- when neutrals stops the ionized particle from 
drifting with the field, and (III) Hall -– when electrons are well coupled or partially coupled to the field and 
ions are partially or completely decoupled from the field. Since Hall parameter $\beta_j$ will vary with the altitude, we shall anticipate that solar atmosphere will be in different non-ideal MHD regimes at different altitudes.

As noted above, non-ideal MHD effects are dependent upon the ambient physical parameters of the medium and 
their relative importance, and consequently, their role in the excitation and propagation of the waves can be 
gauged by the plasma-Hall parameter. In table 1, we give some relevant parameters for the solar photosphere \citep{VAL81} and 
corresponding plasma--Hall parameter. We have assumed that all ions are hydrogen ions although metallic ions 
generally dominate the cold photosphere. However, numbers for metal ions are speculative as the cross-section 
is not known. Assuming cross section $\sigma = 30 \times \sigma_{in}$ for the metallic ions ($m_i = 30\,m_p$, here $m_p = 1.67\,10^{-24}\,\mbox{g}$ is proton mass) gives similar Hall beta parameters as for the ions (table 1). Therefore, we shall refrain from giving Hall parameter separately for the metallic ions. As is clear from the values of $\beta_j$ at different 
altitude, the Ohmic dissipation will be dominant process at the surface of the photosphere. With the increasing altitude, Hall diffusion becomes the important mechanism for the flux distribution in the medium. 
\begin{table*}
 \centering
 \begin{minipage}{140mm}
  \caption{\label{tab:table2} The neutral mass density $\rho = m_n\,n_n$, the ratio of ion to neutral mass density, 
the ion and electron collision frequencies and plasma Hall parameters at different heights of the solar atmosphere 
are given in the table. The ion mass $m_i = m_p$ and the magnetic field $B = 10\,\mbox{G}$ has been assumed.}
  \begin{tabular}{|@{}llrrrrrrlr|@{}|}
  %\begin{tabular}{|llrrrrrlr|}
  \toprule[0.12em]
h\,(km) & $n_n(\mbox{cm}^{-3})$  & $n_e/n_n $ & $\nu_{in}\,(\mbox{Hz})$ & $\nu_{ii}\,(\mbox{Hz})$ &
$\nu_{en}\,(\mbox{Hz})$&$\nu_{ei}\,(\mbox{Hz})$ & $\beta_i $ & $\beta_e $ \\
\midrule[0.12em]
 $0$ & $1.2\cdot10^{17}$ & $5.3\cdot10^{-4}$ & $1.2\cdot 10^{9}$& $2.2\cdot 10^{7}$ &$3.6\cdot 10^{10}$ & $1.3\cdot 10^{9}$ & $10^{-5}$ & $10^{-2}$&   \\
$515$ & $2.1\cdot10^{15}$ & $1.2\cdot10^{-4}$ & $1.8\cdot 10^{7}$ & $2.1\cdot 10^{5}$ &$5.2\cdot 10^{8}$ &  
$1.3\cdot 10^{7}$& $10^{-3}$ & $10^0$ & \\
$1065$ & $1.7\cdot 10^{13}$ & $10^{-2}$ & $1.7\cdot 10^{5}$ & $5.1\cdot 10^{4}$&$5.1\cdot 10^{6}$ & $3.1\cdot 10^{6}$& $10^{-1}$ & $10$ & \\
$1515$ & $\quad\quad10^{12}$ & $6\cdot10^{-2}$ & $1.1\cdot 10^{4}$ & $3.3\cdot 10^{4}$&$3.2\cdot 10^{5}$ & $2.0\,\cdot 10^{6}$&$10^{0}$ & $10^2$ & \\
$2050$ & $7.7\cdot 10^{10}$ & $5\cdot10^{-1}$ & $8.8\cdot 10^{3}$ & $1.6\cdot 10^{4}$&$2.6\cdot 10^{4}$ & $9.6\cdot 10^{5}$&$10^{0}$ & $10^2$ & \\
$2543$ & ${}\quad\quad10^{9}$ & $1.2\cdot 10^{0}$ & $0.9\cdot 10^{2}$ & $1.1\cdot 10^{0}$&$2.6\cdot 10^{3}$ & $6.6\cdot 10^{1}$&$10$ & $10^5$ & \\
\bottomrule[0.12em]
\end{tabular}
\end{minipage}
\end{table*}
Although numbers are not given for $250 \mbox{km}$, Hall starts to competes with the Ohmic diffusion at 
this altitude. For a $10^3\,\mbox{G}$ field, Hall will dominate the Ohmic diffusion at the surface. The ambipolar diffusion is unimportant in the lower photosphere ($ \leq 1500\,\mbox{km}$) if $B = 10\,G$, since $\beta_i \ll 1$. Only at higher altitude, typically between 
$\left(1500 - 2500\right) \,\mbox{km}$, ambipolar diffusion becomes important although Hall is only half as small as ambipolar in this region. Since neutral number density plummets rapidly beyond $2500 \,\mbox{km}$, the role of ambipolar diffusion diminishes in the upper chromosphere. However, since cause of the Hall effect is due to the symmetry breaking between electrons and ions with respect to the magnetic field \citep{PW07}, Hall will continue operating at this height as well albeit on a smaller scale. This conclusion appears counterintuitive since {\it Hall MHD in a fully ionized plasma} generally operates when the frequencies of interest is larger than the ion-cyclotron frequency over ion-inertial scale $\delta_i \equiv v_{Ai} /\omega_{ci}$, a ratio of the ion- \alf speed ($v_{Ai} = B / \sqrt{4\,\pi\,\rho_i}$, $\rho_i = m_i\,n_i$ is the ion mass density) to the ion-cyclotron frequency. For a $10\,G$ magnetic field, the cyclotron frequency is $10^5\,, 10^6$ and $10^7\,,\mbox{s}^{-1}$ at $h = 0, 515$ and $1065\, \mbox{km}$ respectively. Further the Hall scale, in a fully ionized plasma, $\delta_i \approx 2, 44$ and $71\,\mbox{cm}$, corresponding to to $h = 0\,,$ $515\,,$ and $1065\,\mbox{km}$ respectively, is very small. Clearly, the {\it spatial and temporal scales of Hall MHD derived from a fully ionized, two component electron – ion plasma suggest that  the Hall effect is unimportant}. Therefore, fully ionized Hall MHD description of solar atmospheric plasma is irrelevant. Understandably, the role of Hall MHD of a fully ionized plasma in the solar atmosphere has been overlooked by the solar community.

The importance of Hall diffusion in angular momentum transport in protoplanetary discs \citep{W99, BT01, ss1, ss2, sw1, sw2, PW06} indicates that Hall operates on different spatial and temporal scales in a partially ionized plasmas. Recently, \cite{PW07} developed a single fluid formulation for the partially ionized plasma and showed that {\it the spatio-temporal scale over which  Hall operates in a partially ionized plasma is very different from a fully-ionized Hall-MHD}. Defining Hall frequency as \citep{PW07}
\bq
\omega_H = \frac{\rho_i}{\rho}\,\omega_{ci}\,,
\eq
 where $\rho = \rho_n + \rho_i = m_n\,n_n + m_i\,n_i$ is the mass density of the bulk fluid with $m_n\,, m_i$ as the mass and $n_n\,, n_i$ as the number density of the neutral and ion fluids respectively, we can demand that if dynamical frequency of the system is larger than the Hall frequency, i.e. $\omega_H \lesssim \omega$, then Hall effect will be important. Furthermore, Hall in a partially ionized medium will not operate on the ion-inertial scale as in a fully ionized plasma, but over \citep{PW07} 
\bq
L_H = \left(1 + \frac{\rho_n}{\rho_i}\right)^{\left(1/2\right)}\,\delta_i\,.
\eq
Assuming $m_i = m_n$, we see from Table 1 that 
Hall frequency varies between $10 –- 10^5\,\mbox{Hz}$ and thus for fluctuations occurring at higher than $\omega_H$, Hall will play an important role over $L_H \sim$ few hundred meters to few kms in the medium. We note that unlike high frequency \alf wave which damps in the solar medium \citep{LAK05}, the excitation of low-frequency normal modes due to collisional coupling will propagate un-damped in the medium. 

We investigate the propagation of waves in the solar photosphere in the Hall regime using recently developed general 
set of equations by \cite{PW07} applicable to the partially ionized plasmas. The effect of dissipative diffusion on 
the wave properties has been investigated in the past \citep{TM62, KP69}. The wave dissipation in such a medium is 
dependent not only on the ion-neutral collision frequency but also on the fractional ionization of the 
medium \citep{KR03, PW07}. The wave damping have recently been studied in the context of spicules 
dynamics \citep{BDP}. In the lower solar photosphere however, since Hall will dominate all other diffusive processes 
the resultant low frequency ion-cyclotron and collisional whistler wave will be the normal mode of the medium \citep{PW07}. 

We investigate the linear and nonlinear wave properties of the medium in the present work. We show that since such a medium is 
inherently dispersive, the balance between the dispersion and nonlinearity leads to DNLS equation. The basic set 
of equations and the linearized dispersion relation is discussed in Sec. II . In Sec. III we first discuss 
the parametric instability and then describe the nonlinear equation. In Sec. IV discussion of the results and a brief 
summary is presented.

\section{Basic model}
The solar photosphere is a weakly ionized medium consisting of electrons, ions, neutrals. The single fluid description of such a plasma has been given in the past \citep{C57, BR65}. We shall use the single fluid description given by \cite{PW07}.

The continuity equation for the bulk fluid is given as
\bq
\frac{\partial \rho}{\partial t} + \grad\cdot\left(\rho\,\v\right) = 0\,,
\label{eq:cont}
\eq
where $\rho = \rho_e + \rho_i + \rho_n \approx \rho_i + \rho_n$ is the bulk fluid density and $\v$ is the bulk 
velocity, $ \v = (\rho_i\,\vi + \rho_n\,\vn)/\rho$ with $\rho_i,\,\vi$ and $\rho_n,\,\vn$ as the mass density and bulk velocities 
of the ion and neutral fluids respectively. The momentum equation is given as
\bq
\rho\,\frac{d\v}{dt}=  - \nabla\,P + \frac{\J\cross\B}{c}
\label{eq:meq}.
\eq
Here $\J = n_e\,\left(\vi - \ve\right)$ is the current density, $\B$ is the magnetic field and $P = P_e + P_i + P_n$ is the total pressure.
The induction equation is
\begin{eqnarray}
\delt \B = \curl\left[
\left(\v\cross\B\right) - \frac{4\,\pi\,\eta}{c}\,\J - \frac{4\,\pi\,\eta_H}{c}\,\J\cross\hB
\right. \nonumber\\
\left.
+ \frac{4\,\pi\eta_A}{c}\,
\left(\J\cross\hB\right)\cross\hB
\right]\,,
\label{eq:ind}
\end{eqnarray}
where $\hB = \B /B$, and the Ohmic ($\eta$), ambipolar ($\eta_A$) and
Hall ($\eta_H$) diffusivity are 
\bq \eta =
\frac{c^2}{4\,\pi\sigma}\,\,, \eta_{A} =
\frac{D^2\,B^2}{4\,\pi\,\rho_i\,\nu_{in}}\,, \eta_H =
\frac{c\,B}{4\,\pi\,e\,n_e}\,.
\label{eq:diffu}
 \eq
Here $D = \rho_n/\rho$. We see from above Eq.~(\ref{eq:ind}) that the ratio of the Hall $(H)$ and the Ohm $(O)$ 
gives $H/O \sim \beta_e$ and the ratio between ambipolar $(A)$ and Hall $(H)$ gives $A/H \sim D^2 \,\beta_i$. In 
the photosphere, $D \rightarrow 1$, $\beta_i \ll 1$ and thus ambipolar diffusion can be neglected. In Fig.1, we show the ratio of $A/H$ and $H/O$ for the collision frequencies from table 1 for $B = 100\,G$ field. The figure suggest that between $0 \le h \le 620\,\mbox{km}$ in the solar atmosphere, Ohm will dominate both non dissipative Hall and dissipative ambipolar diffusion; between $620\,\mbox{km} \le h \le 1160\,\mbox{km}$ Hall dominates Ohm as well as ambipolar whereas between $1160\,\mbox{km} \leq h \leq 2500 \,\mbox{km}$ ambipolar will dominate Hall. Clearly these estimates are based on the  
\begin{figure}
\includegraphics[scale=0.38]{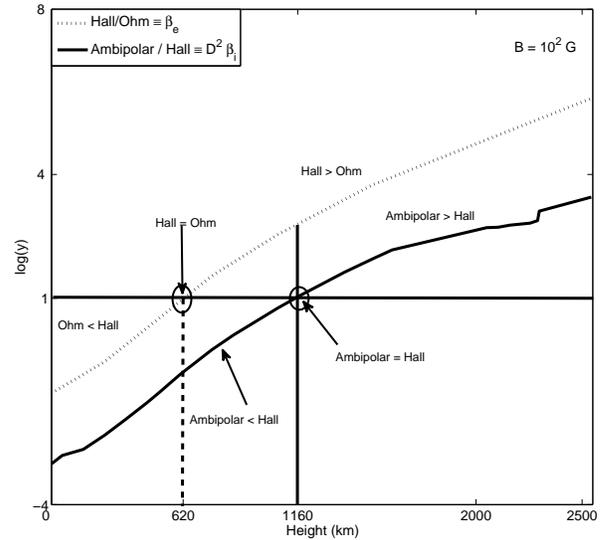}
\caption{\large{The ratio of Hall to Ohm ($\beta_e$), and ambipolar to Hall ($D^2\,\beta_i$) is given for a $B = 10^2\, G$ field.}}
\end{figure}
constant magnetic field strength. In Fig.~2, we show that ambipolar to Hall ratio changes with the changing field strength. For example, if $B = 10\,G$, the Hall will dominate the ambipolar in  $1000\,\mbox{km} \leq h \leq 1600 \,\mbox{km}$ and only after $1600\,\mbox{km}$ ambipolar becomes more important than Hall. Below $100\,\mbox{km}$ Ohmic dissipation dominates all other form of diffusion. With increasing magnetic field strength, the domain of Ohmic dissipation shrinks and Hall starts operating at much closer to the surface. Therefore, we may say that for sufficiently strong field $B > 1\,\mbox{kG}$, Hall diffusion will operate in the photosphere whereas ambipolar diffusion will 
operate in the chromosphere.
\begin{figure}
\includegraphics[scale=0.38]{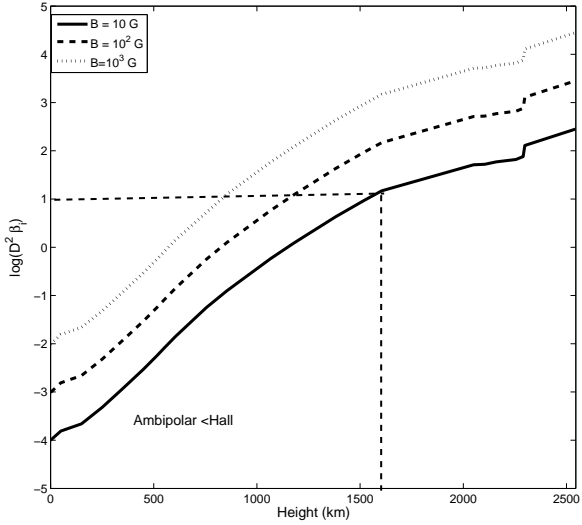}
\caption{\large{The ratio of ambipolar and Hall term is shown in this figure for varying magnetic field strength.}}
\end{figure}
Since Hall scale is dependent upon the fractional ionization of the medium \citep{PW07}, with increasing altitude, the Hall scale 
$L_H = \eta_H/ v_A $ will shrink and finally approach the ion inertial scale for a fully ionized plasma.

The role of collisional effects in the photosphere and chromosphere has been investigated in the past \citep{G04} using the conductivity tensor approach. 
We note that in the conductivity tensor approach plasma inertia is ignored while deriving generalized Ohm's law, and MHD waves (which require non-zero plasma inertia!) are studied using such a generalized Ohm's law. Clearly to overcome this logical inconsistency, a general set of equations Eqs.~(\ref{eq:cont}), (\ref{eq:meq}), and (\ref{eq:ind}) have been derived by \cite{PW07}. These set of equations can be closed by prescribing a thermodynamic relation between pressure and mass density. In order to compare the results of present formulation with the previous investigations e.g. \citep{G04}, we note that the generalized Ohm's law $\J = \sigma_{\parallel}\, \E_{\parallel}' + \sigma_{\perp}\, \E_{\perp}' + \sigma_H\,\E'\cross \B/B$ (here $\E' = \E + \v \cross \B$ is the field in the bulk frame, $\E_{\parallel} = \E\cdot\hB$, $E_{\perp} = \hB\cross\left(\E\cross\hB\right)$ where $\hB = \B/B$) can be inverted to yield
\bq
\E' = \frac{\J}{\sigma_{\parallel}} - \left( 
\frac{\sigma_P}{\sigma_{\perp}^2} - \frac{1}{\sigma_{\parallel}}
\right)\,\J_{\perp} - \frac{\,\sigma_H}{\sigma_{\perp}^2}\,\left(\J\cross\hB\right)\,. 
\label{eq:indx} 
\eq
Here $\sigma_{\perp} = \sqrt{\sigma_{P}^2 + \sigma_{H}^2}$ and $\J_{\perp} = \hB\cross(\J\cross\hB)$. The parallel ($\sigma_{\parallel}$), Hall ($\sigma_H$) and Pedersen ($\sigma_{P}$) conductivities are given by \cite{WN99}
\begin{flushleft}
\bq
\sigma_{\parallel} = \frac{c\,e\,n_i}{B} \left(\beta_e + \beta_i\right)\,,
\label{eq:cPL}
\eq
\bq
\sigma_{P} = \frac{c\,e\,n_i}{B} \left(\frac{\beta_e}{1 + \beta_e^2} + \frac{\beta_i}{1 + \beta_i^2}\right)\,,
\label{eq:cP}
\eq
\bq
 \sigma_{H} = \frac{c\,e\,n_i}{B}\,\left(
- \frac{\beta_e^2}{1 + \beta_e^2} + \frac{\beta_i^2}{1 + \beta_i^2}\right)\,.
\label{eq:cH}
\eq
\end{flushleft}
While writing above expressions for the conductivity tensors, we have assumed that plasma is quasineutral, i.e. $n_e \approx n_i$. Taking curl of Eq.~(\ref{eq:indx}) and using $c\,\curl \E' = - \partial_t \B$ we get the induction equation (\ref{eq:ind}) except that $\eta\,, \eta_H$ and $\eta_A$ will be expressed as a combination of $\sigma_{\parallel}, \sigma_P$ and $\sigma_H$. Assuming $\beta_e \gg 1$ and $\beta_i \sim 1$, we note that 
\bq
\hat{\sigma}_{\parallel} \approx \beta_e\,,
\hat{\sigma}_{P} \approx \beta_i\,,
\hat{\sigma}_{H} \approx 1\,,
\eq
where $\hat{\sigma} = B\,\sigma / \left(c\,e\,n\right)$. In the $\beta_e \gg 1$ and $\beta_i \sim 1$ limit , the ratio of ambipolar and Hall terms from Eq.~(\ref{eq:indx}) can be written as 
\bq 
\frac{\left( 
\frac{\sigma_P}{\sigma_{\perp}^2} - \frac{1}{\sigma_{\parallel}}
\right)}{ \frac{\,\sigma_H}{\sigma_{\perp}^2}} \sim \beta_i\,,
\eq
which is same as derived from induction Eq.~(\ref{eq:ind}) except for the $D^2$ factor. This difference arises because in the conductivity tensor approach, ion inertia is set to zero and thus $D \equiv \rho_n /\left(\rho_i + \rho_n \right) = 1$ since $\rho_i = 0$. Clearly the induction equation of \cite{PW07} is more general than the conductivity tensor approach which is valid only for weakly ionized plasma and is ill suited to describe the transition regions where plasma inertia is important. The ratio of Hall to Ohm from Eq.~(\ref{eq:indx}) becomes
\bq
\sigma_{\parallel}\,\frac{\sigma_H}{\sigma_{\perp}^2} \sim \beta_e\,,
\eq      
which is the same as derived from eq.~(\ref{eq:ind}). Clearly in the weakly ionized limit \cite{PW07} and conductivity approach gives identical results. It has been suggested in the past that Pedersen current dissipation could play an important role in the chromosphere \citep{G04}. We see from Fig.~ 2 that ambipolar and Hall effect dominates the lower chromosphere (particularly for a strong magnetic field) and thus, the dissipative Pedersen current heating could be a plausible mechanism for coronal heating.   

Eqs.~(\ref{eq:cont}), (\ref{eq:meq}), and (\ref{eq:ind}) together with an equation of state $P = c_s^2\,\rho$ forms a closed set. 
We note that the collisional dissipation will cause the loss of energy in the lower solar atmosphere and therefore, generally a proper energy equation should be used for a more realistic modelling of the physical processes \citep{AHL07}. However, in order to keep the description simple, we shall use $P = c_s^2\,\rho$ to investigate the wave properties of the medium. In order to study waves in the photosphere, we shall assume a homogeneous background with no flow ($\v_0 = 0$). The equations (\ref{eq:cont}),(\ref{eq:meq}) 
and (\ref{eq:ind}) after linearizing around $\B_0,\, P_0\,\rho_0$ and Fourier analysing 
$exp \left(i\,\omega\,t -– i\, \k\cdot\x\right)$ becomes
\begin{equation}
\omega\,\drho - \rho\,\k\cdot\dv = 0\,,\nonumber\\
\end{equation}
\begin{equation}
\omega\,\dv = \frac{-1}{4\,\pi\,\rho} \left[\left(\k\cdot\B\right)\dB -
\left(\frac{\omega^2}{\bar{\omega}^2}\right)\left(\dB\cdot\B\right)\,\k \right].
\end{equation}
Here $\bar{\omega}^2 = \omega^2 - k^2\,c_s^2$.
Defining $\kh\cdot\hB = \cs$, along with \alf frequency $\omega_A = k\,v_A$, for $\dB$ we get 
\begin{eqnarray}
\left(\omega^2 - \omega_A^{2}\,\cos^2\theta\right)\dB =  
 \frac{\omega^2}{{\bar{\omega}}^2}\,\omega_A^2 \left(\dB\cdot\hB\right)\,\left(\hB - \kh\,\cs \right)
\nonumber \\
 - i\, k^2\,\eta_H\,\omega\,\cs\,\left(\hat{\k}\cross\dB\right)\,.
\label{eq:Lind}
\end{eqnarray}
Only Hall and convective terms have been retained while linearizing the induction Eq.~(\ref{eq:ind}). After some 
straightforward algebra, following dispersion relation can be derived from equation (\ref{eq:Lind})
\begin{eqnarray}
\left( \omega^2 - \omega_A^{2}\,\cos^2\theta  \right)\times
\nonumber\\
\,\left( \omega^2\left(1 - \frac{\omega_A^2}{\bar{\omega}^2}\,\sin^2\theta\right) - \omega_A^{2}\,\cos^2\theta  \right) 
= k^4\,\eta_H^2\,\omega^2\,\cos^2\theta \,.
\label{eq:DER}
\end{eqnarray}

Since compressional mode will not affect the transverse mode, we shall drop $k^2\,c_s^2$ term in the above 
dispersion relation. Then writing $ k^2\,\eta_H = \omega_A^2/\omega_H$, we can recast above Eq.~(\ref{eq:DER}) as
\bq
\omega^2   =  \omega_A^2 \cos^2\theta \pm \,\left(\frac{\omega}{\omega_{H}}\right)\,\omega_A^2\,\cs\,.
\label{hal2}
\eq
This dispersion relation acquires a familiar form (cf.\ Wardle \& Ng 1999, Eq.~25) when wave is propagating along 
the ambient magnetic field ($\theta = 0$)

In the $\omega_A \ll \omega_{H}$ limit, i.e. when the 'dressed ions' (with effective mass $m^* = \rho/n_i$) gyrates faster than the time over which magnetic fluctuation propagates, the dispersion relation (\ref{eq:DER}) gives the familiar Alfven wave 
\bq
\omega^2 \simeq \omega_A^2\,\cos^2\theta\,.
\label{alf}
\eq
We note that these waves propagate in the neutral medium. The propagation of these low frequency \alf waves in a neutral medium occurs due to collisional coupling of the plasma particles with neutrals. The electrons, ions and neutrals are well coupled by the collisions and partake in the oscillations together. 

The propagation of waves in a collision dominated medium appears counter intuitive since one would expect that collision will 
dissipate the energy and thus damping of the fluctuations will occur. This picture is valid only when wave frequencies are comparable or greater than the 'effective' neutral-ion collision frequency, i.e. $k\,v_A \gtrsim  \sqrt{\rho_n/\rho_i}\,\left(\beta_e/1 + D\,\beta_e\right)\,\nu_{ni}$ implying that 
the neutral is not hit often enough by the ions to participate in the oscillations \citep{PW07}. Therefore, only fluctuations of certain 
wavelength will disappear due to collisional dissipation, and waves of wavelengths $\lambda$ exceeding 
$\lambda_{\mbox{cutoff}} = \sqrt{2}\,\pi\,D\,v_A/\nu_{ni}$ can propagate in the medium. Since magnetic restoring force 
due to field deformation of wavelength $\lambda$ acts only on the plasma particles, the neutrals at a 
distance $\lambda$ apart can respond simultaneously to this restoring force only if the communication time 
between the neutrals ($\sim \lambda /v_A$) is smaller than the 'effective' collision time $t_{c} \sim D\,\nu_{ni}^{-1}$. Only when 
$t_{c}$ exceeds $\sim \lambda /v_A$, the wave in the medium will damp. Using table 1, we see that 
that since $\lambda_{\mbox{cutoff}}$ is $0.01\,\mbox{cm}$ at $h = 0$ and $7\,\mbox{m}$ at $h = 1000\,\mbox{km}$, 
the low frequency \alf wave of large wavelengths will propagate in the partially ionized solar photosphere without damping. 
Therefore we can say that the solar photosphere supports the excitation and propagation of the low frequency \alf waves in 
the predominantly {\it neutral} medium where the inertia of the fluid is carried by the neutral fluid and the 
magnetic deformation is felt by the plasma particles.        

When $\omega_{H}\ll \omega_A$, i.e. when the 'dressed ion' gyration period is slower than the propagation time of the magnetic fluctuations in the medium, the 
dispersion relation (\ref{eq:DER}) can be analysed in low $\omega \ll \omega_A$ and high $\omega_A \ll \omega$ frequency limits. In the low frequency limit, we get 
\bq
\omega^2 \simeq  \omega_{H}^2\,\cos^2\theta \,.
\label{eq:es}
\eq
In the high frequency $\omega_A \ll \omega$ limit,
\bq
\omega^2 \simeq  \left(\frac{\omega_{A}^2}{\omega_{H}}\right)^2\,\cos^2\theta\,.
\label{eq:em}
\eq
The mode described by Eq.~(\ref{eq:es}) is the modified electrostatic ion cyclotron mode. Since $\omega \simeq \omega_H \sim 
10 –- 10^5\,\mbox{Hz}$, solar photosphere can support ion cyclotron waves, except when the direction of wave propagation is almost transverse to the ambient magnetic field. Equation (\ref{eq:em}) is the dispersion relation 
for the whistler mode. In the lower photosphere, the whistler is high frequency branch whereas ion-cyclotron frequency is the low frequency branch of the normal mode of the medium.

\section{The parametric instability and the solitons}
Assuming a uniform background field in the z-direction and the variation of the field along this direction only,
above set of Eqs.~ (\ref{eq:cont}), (\ref{eq:meq}) and (\ref{eq:ind}) can be written as
\begin{eqnarray}
\dt{\rho}  + \dpz{\left(\rho\,v_z\right)} = 0\,,
\nonumber\\
\left(\dt{} + \vz\,\dpz{} \right)\,\vz = - \frac{1}{\rho}\dpz{}
\left(p + \frac{|B|^2}{8\,\pi}\right),
\nonumber\\
\left(\dt{} + \vz\,\dpz{} \right)\,v = \frac{\Bz}{4\,\pi\,\rho}\,\dpz{B},
\nonumber\\
\dt{B} = - \dpz{}\left(\vz\,B - \Bz\,v +\frac{i\, C_I\delta_H\,\rho_0}{\rho}
\,\dpz{B} \right) \,.
\label{eq:cmp1}
\end{eqnarray}
Here  $C_I = \Bz/ \sqrt{4\,\pi\,\rho_0}$, $\delta_H = v_A/\omega_H$ is the neutral skin depth, $C_I\,\delta_H = \Bz\,\eta_H / B$,  
$B = B_x + i\,B_y$ and $v = v_x + i\,v_y$. A medium described by the above set of Eqs.~(\ref{eq:cmp1})
admits circularly polarized linear waves $(B, v) = (B_0,\, U_0)\,\exp{i\,\left(\omega_0\,t -– k_0\,z\right)}\,$, as
an exact solution. This can easily be seen by linearizing Eqs.~(\ref{eq:cmp1}) around a static equilibrium
state. The linear dispersion relation provides a relation between the wave number $k_0$ and frequency $\omega_0$ of the
wave
\bq
\omega_0^2 = k_0^2\,C_I^2\,\left(1 \pm
\frac{\omega_0}{\omega_{H}}\right)\,.
\label{eq:pdr}
\eq
The amplitudes $B_0$ and $U_0$ of the waves are related by
\bq
U_0 = - \left(\frac{B_0\,\omega_0}{k_0\,B_z}\right)\left(1 \pm
\frac{\omega_0}{\omega_{H}}\right)^{-1}\,.
\eq
We note that the stability analysis of the steady-state background consisting of the unperturbed as well as
the circularly polarized pump waves, i.e. $\B = (B_x(z), B_y(z), B_z)$ and $\u = (U_x(z), U_y(z), 0)$ with a
constant density, indicates that the system is unstable, the so called modulational instability. The modulational
instability has been extensively studied in the past \citep{H74, S76, D78, G78, BS88, PV07}. Assuming such a steady-state
background and linearizing  Eqs.~(\ref{eq:cmp1}) one ends up with the following set of equations.
\begin{eqnarray}
\delt\drho + \rho\,\delz\duz = 0\,,
\nonumber\\
\delt\dup - \omega_0\,\dum = \frac{B_z}{4\,\pi\,\rho}\,
\left( \delz\dBpl + k_0\,\dBm\right)\,,
\nonumber\\
\delt \dum + \omega_0\,\dup - k_0\,U_0\,\duz = \nonumber\\
 \frac{B_z}{4\,\pi\,\rho}\,
\left( \delz \dBm - k_0\,\dBpl\right) + k_0 \, \frac{B_z\,B_0}{4\,\pi\,\rho^2}\,\drho\,,
\nonumber\\
\delt \duz = - \frac{C_s^2}{\rho} \,\delz\drho
- \frac{B_0}{4\,\pi\,\rho}\,\delz\dBpl\,,
\nonumber\\
\delt\dBpl - \omega_0\,\dBm = B_z\left( \delz\dup + k_0\,\dum\right)
- B_0\,\delz\duz \nonumber\\
- \eh \left( \frac{\partial^2 \dBm}{\partial z^2}
- 2\,k_0\,\delz \dBpl + k_0^2\,\dBm\right)
 -\frac{\eh\,k_0\,B_0}{\rho}\delz\drho\,,
\nonumber\\
\delt\dBm + \omega_0\,\dBpl = B_z\left( \delz\dum - k_0\,\dup\right)+ k_0\,B_0\,\duz\
\nonumber\\
+ \eh \left( \frac{\partial^2 \dBpl}{\partial z^2}
+ 2\,k_0\,\delz \dBm - k_0^2\,\dBpl\right)\,,
\label{eq:lfin}
\end{eqnarray}
While writing the above set of equations, we have assumed that $\dBz = 0$. Here
$\dBpl = \dBx\,\cos\phi + \dBy\,\sin\phi\,$, $\dBm = \dBx\,\sin\phi
- \dBy\,\cos\phi\,$, $\dup = \dux\,\cos\phi + \duy\,\sin\phi\,$, $\dum = \dux\,\sin\phi
- \duy\,\cos\phi\,$, and $\eh = B_z \,\eta_H/ B$. Fourier analyzing the above set of equations (\ref{eq:lfin}),
we will end up with following $8^{th}$ order dispersion relation.
\bq
\omega^8 + a_7\, \omega^7 + \ldots + a_1\,\omega + a_0 = 0\,.
\label{eq:dr}
\eq
The coefficients $a_7,\ldots a_0$ are given in the Appendix. Defining
$\omega = \omega / \omega_0$, $x = k/k_0$, $b = \omega_0/ \omega_H$, $F_{\pm} = 1/\left( 1 \pm b\right)$,
$\beta = C_s^2/C_I^2$, $k^2\,C_s^2 = \omega_0^2\,x^2\,\beta\, F$, $a = B_0/ B_z$ and noting that since the transition from stability to instability proceeds through
$\omega = 0$, in the $\omega \rightarrow 0$ limit, we see that
\bq
\omega = - \frac{a_0}{a_1}\,.
\label{eq:ldr}
\eq
In the long wavelength limit retaining only $\sim O(k)$ terms in the coefficients $a_1$ and $a_0$ one may write
\bq
\frac{a_1}{F_{\pm}^2} = a^2\,\left[F_{\pm}^{-1}  - \left(1 + b\right)\right] -2\,F_{\pm}^{-1}\,x 
\,,
\eq
\bq
\frac{a_0}{F_{\pm}^2} = - \left(F_{\pm}^{-2} + a^2 \right)\left[1 - F_{\pm}\,\left(b + 1\right) \right]\,. 
\label{eq:caz}
\eq
Since for the left-circularly polarized pump waves, $F_{+} = 1/\left(1 + b\right)$, from Eq. (\ref{eq:caz}),
$a_0 = 0$. Thus in view of Eq. (\ref{eq:ldr}), we shall anticipate that the instability should decreases in the vicinity of $k = 0$. The numerical solution of the dispersion relation (\ref{eq:dr}) for the left circularly polarized (Fig. 3(a)) indeed displays this feature. The normalized growth rate is plotted against the normalized wave number for 
$\omega_0/\omega_{H} = 0.01, 0.1$ and $0.15$. We see that the growth rate decreases in the vicinity of $k \rightarrow 0$. With the increase in
$\omega_0/\omega_{H}$, the growth rate increases. As is clear from the
wave dispersion relation (\ref{eq:pdr}), the increase in $\omega_0/\omega_{H}$ implies the increasing importance of the
Hall term ($\J\times\B$), which appears due to the relative drift between the plasma and the neutral (i.e. ion
Hall-$\beta_i \le 1$). This drift is caused entirely by the collisional coupling of the ions with the neutrals. Therefore,
the increase in $\omega_0/\omega_{H}$ implies the enhanced availability of the free energy reservoir. Hence with the
increasing $\omega_0/\omega_{H}$, growth rate increases.

\begin{figure}
\includegraphics[scale=0.38]{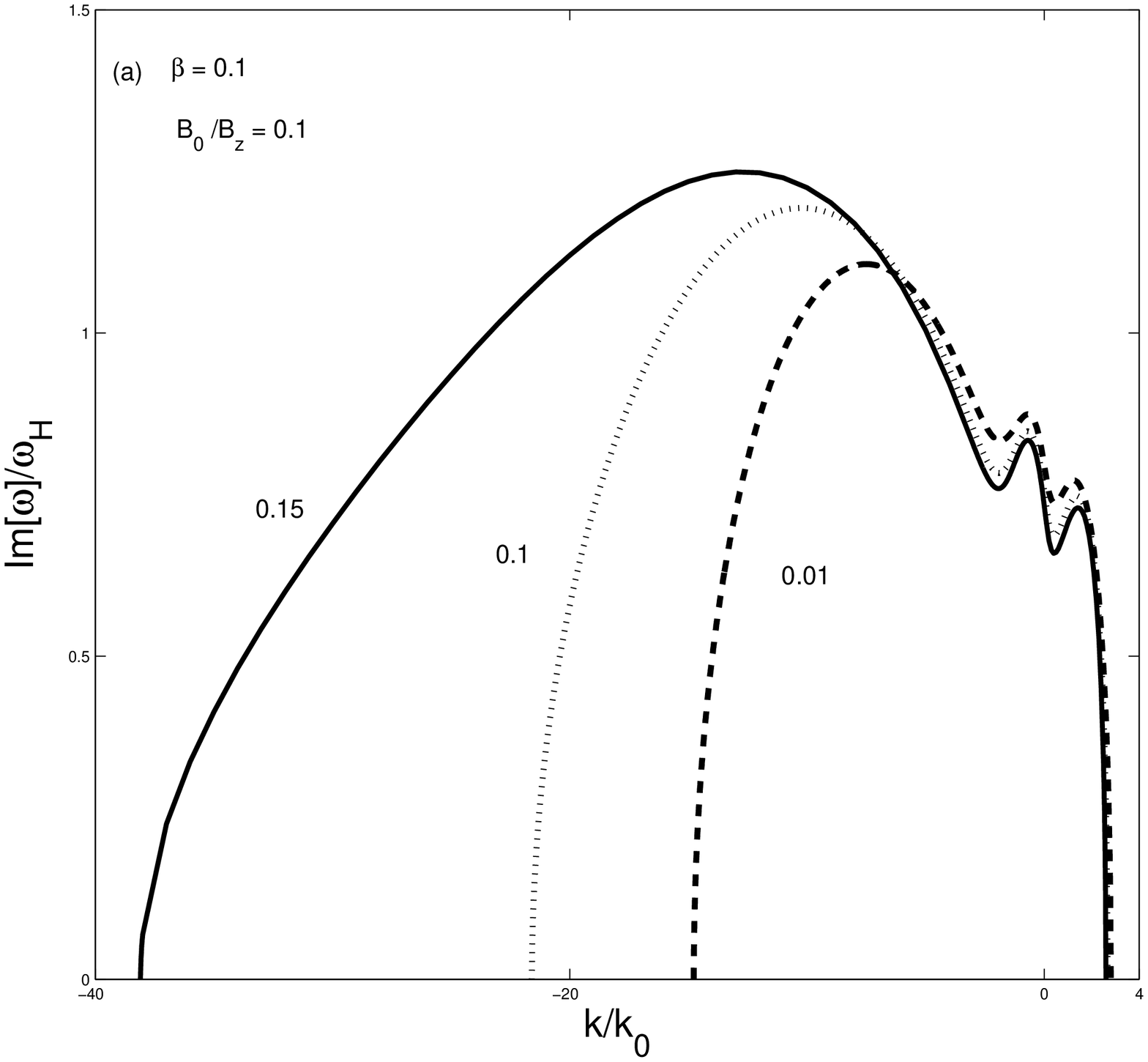}
\includegraphics[scale=0.38]{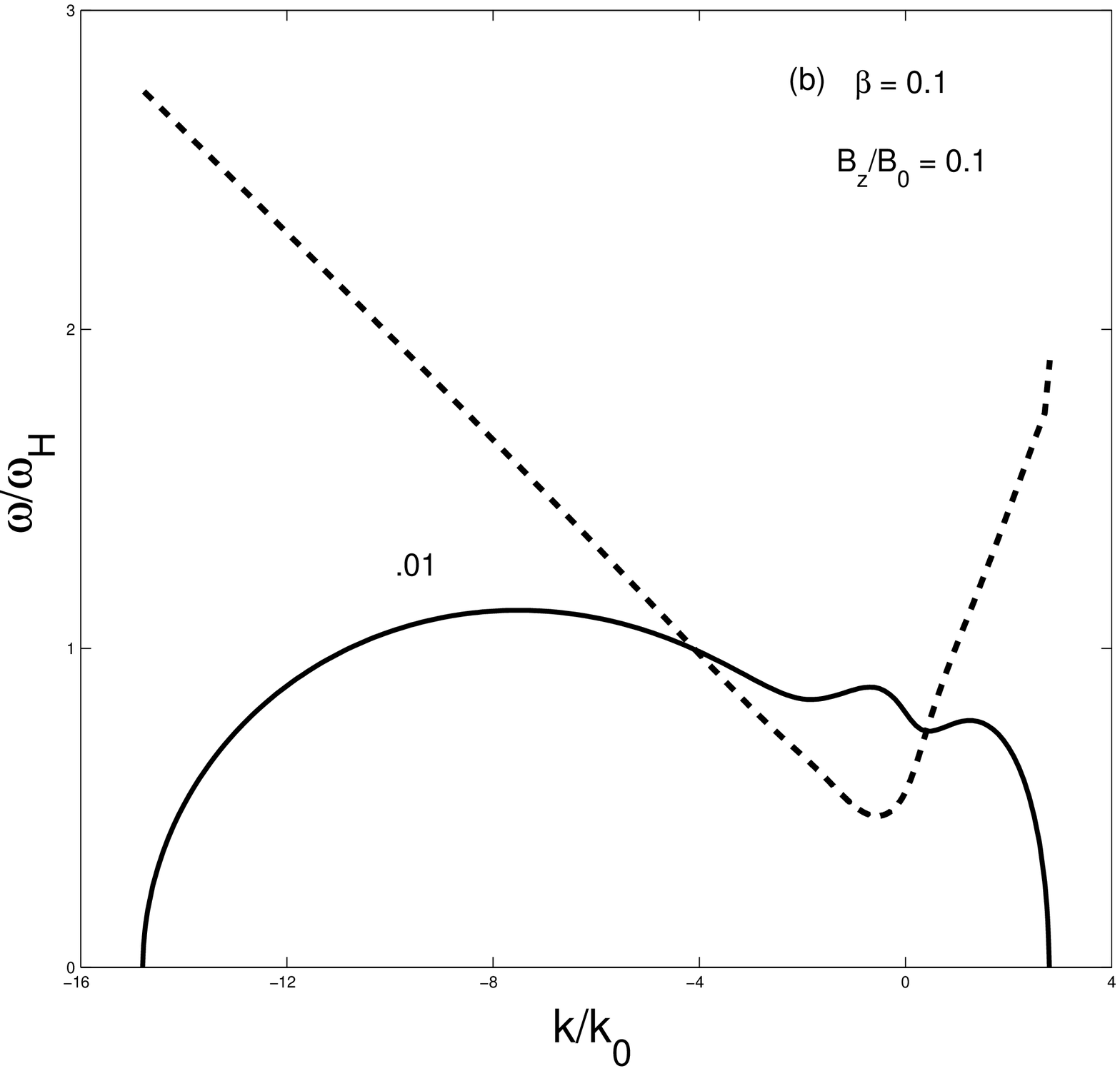}
\caption{\large{The growth rate of the left
circularly polarized wave $Im[\omega]/\omega_H$ against $k/k_0$ is
shown in Fig. 3(a) for $\omega_0/\omega_H = 0.01\,, 0.1$ and $0.15$ with plasma $\beta = 0.1$ and $B_0/B_z = 0.1$ . In Fig. 3(b) both real (dotted line) and imaginary (solid line) part of the frequency is given for $\omega_0/\omega_H = 0.01$. All other parameters are same as in Fig.~3(a).}}
\end{figure}
In Fig.~3(b) we plot the real (dotted line) and imaginary (solid line) part of the frequencies for $\omega_0/\omega_H = 0.01$. We see from Fig.~3(b) that the growth rate of instability causes the sharp decrease of the real frequency. It suggests that most of the available free energy of the pump is invested in the growth of the fluctuations resulting in the decay of the real part of the frequency. However, with the decrease in the growth rate, most of the the pump energy goes to the real part of frequency, as seen from the figure. Further, we note that before the real part of frequency shows signs of recovery, the imaginary part start to fluctuate and finally drops to zero.    

In Fig.~4 the growth rate for the left circularly polarized wave is plotted against the wave number, for different values of plasma beta, $\beta = 0.01\,, 1\,, \mbox{and}\, 5$. Although the growth rate is dependent upon the plasma compressibility, the dependence is not linear. For example, when plasma $\beta$ increases
from $0.01$ to $5$, the growth rate almost doubles. However, when both acoustic and \alf speed equals, i.e. $\beta = 1$,
the growth rate is highest. For $\beta = 10$, the growth rate is similar to the case when $\beta = 5$. Therefore, the
highest growth rate is when $v_A \approx c_s$. The growth rate is not very sensitive to the the variation of the pump amplitude,
$B_0/ B_z$ and hence is not given here.
\begin{figure}
\includegraphics[scale=0.38]{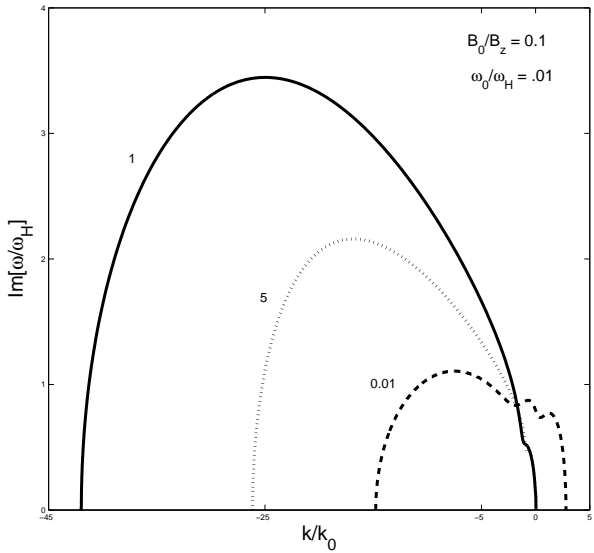}
\caption{\large{Same as in Fig.~1, with varying plasma $\beta$.}}
\end{figure}

 For the right-circularly polarized waves, when $F_{-} = 1/(1 -\omega_0/\omega_{H})$, the waves can grow at a rate
\bq
Im[\omega] \approx F_{-}\,,
\label{eq:lres}
\eq
in the vicinity of $x = 0$.
Since the growth rate (Eq.~\ref{eq:lres}) is inversely proportional to $(1- \omega_0/\omega_{H})$,
near $\omega_0 \simeq \omega_{H}$, when wave frequency matches the Hall frequency, the instability can grow resonantly.
Therefore, the growth of the instability is quite different for the left and right-circularly waves. Whereas for the
left-circularly polarized waves, the instability decreases in the neighbourhood of $k = 0$, for the
right-circularly polarized pump, the instability may become large near $k = 0$. However, for the right circularly
polarized waves, unbounded growth of the instability is not possible since linear approximation will break down
once fluctuation becomes comparable to the background equilibrium quantities.
\begin{figure}
\includegraphics[scale=0.38]{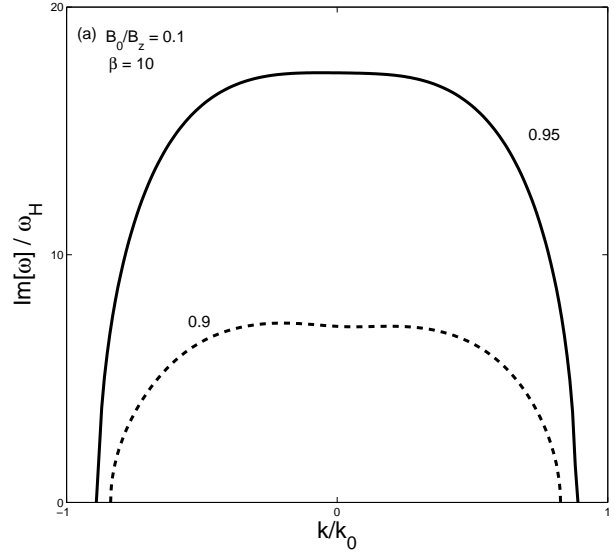}
 \includegraphics[scale=0.38]{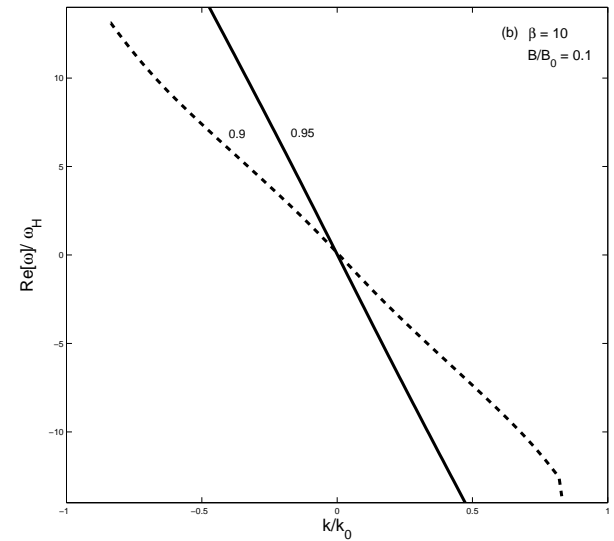}
\caption{\large{The dependence of the growth rate on the amplitude of the
right handed circularly polarized pump wave is shown in Fig.~5(a)
and corresponding real part of the frequency is shown in Fig.~5(b)
for $\omega/\omega_{H} = 0.9$, and $0.95$. The plasma
$\beta = 10$ and $B_0/B_z = 0.1$.}}
\end{figure}

In Fig.~5(a) the growth rate and in Fig. 5(b) corresponding real part
of the frequency is shown for the right-circularly polarized waves.
When $\omega_0/\omega_{H} = 0.9$, the growth rate becomes very
large. The free energy is resonantly pumped into the fluctuations
with increasing $\omega_0/\omega_{H}$. The physical system behaves
more like a driven oscillator. The resonant driving is indirectly related
to the neutral-plasma collisions. The relative drift between the
plasma and the neutral causes a Hall field over the Hall
time $t_H$($t_H \equiv \omega_H^{-1}$). If the \alf wave propagation time $\omega_0^{-1}$ becomes
comparable to the Hall time $t_H$, the energy is freely fed to the fluctuation by the  pump to the plasma particles.
Resulting free energy causes the large growth rate. This behaviour can be seen from the analytical
expression Eq. (\ref{eq:lres}). It should be noted from the corresponding curve in Fig.~5(b) that the real frequency shows a
sharp decline for $\omega_0/\omega_{H} = 0.9$. This suggest that almost all the free
pump energy is resonantly used in the fluctuation growth. Similar behaviour is also noted for
$\omega_0/\omega_{H}= 0.95$. Since the instability growth rate in
this case is larger than when $\omega_0/\omega_{H} = 0.9$, the part of the available free energy
drains faster from the real part of the frequency. This is in agreement with the well known behaviour of the
oscillators near resonance although present system is more complex.

It is believed that the granulations or convective motions are responsible for the \alf wave generation in the photosphere. However the medium is capable of exciting very low frequency fluctuations only, i.e. when the wave frequency is much smaller than the collision frequency. Then the neutrals are 'dragged along' by collisions and as a result, they participate in the collision \citep{TM62}. However, excitation of the high-frequency ideal MHD \alf mode is very unlikely. The high-frequency ideal MHD mode will be damped by collisions of plasma particles with the neutrals as well as collisions of plasma particles with each other \citep{TM62, PVP07, VPP07, VPPD07}.    

The excitation of very low frequency \alf wave is a promising candidate for heating and acceleration of the solar plasma from coronal holes.
Taking $\omega_H = 1\,\mbox{Hz}$ for $\omega_{ci}  \sim 10^4 \mbox{Hz}$ we see that
the growth rate of the left-circularly polarized wave $0.3\, \omega_0$ suggests that the instability of the \alf
wave could be relevant to the turbulent heating. Since $\omega_H = 10 –- 10^5 \mbox{Hz}$, the resonance condition for
the right circularly polarized wave implies that $\omega_0 $ must be in the same range. Therefore, it is quite likely
that the low frequency right circularly polarized waves may resonantly extract energy form the ambient surrounding.

\section{Discussion and summary}
The solar photosphere is a weakly ionized gas with neutral hydrogen carrying the inertia of the fluid. We show that the
non-ideal effects are of paramount importance for the excitation and propagation of the waves. The hydromagnetic waves in
such a medium are an outcome of the balance between the inertia of the neutral component with the deformation of the
magnetic field, to which neutrals are tied only by the ion-neutral collisions. Therefore, very low frequency Alfven wave can be excited in such a medium. The high frequency \alf mode (in the ion fluid) will be damped in such a medium \citep{VPP07, VPPD07}. 

The Hall diffusion dominates the ambipolar and Ohmic diffusion in photosphere. The Hall scale  $L_H = (\rho/\rho_i)^{1/2}\,\delta_i$, is typically two order of magnitude larger than the usual ion-inertial length $\delta_i = v_{Ai}/\omega_{ci}$ \citep{PW07}. The presence of the Hall effect will cause the excitation of the ion-cyclotron and whistler waves in the medium.

We show that in an inhomogeneous photosphere, circularly polarized whistlers, are an exact solution of the ensuing equations. These waves can act as a pump wave. The resulting parametric instability can make the medium turbulent. 

We note that derivative nonlinear Schr$\ddot{\mbox{o}}$dinger equation (DNLS) equations can be derived from  Eqs.~(\ref{eq:cmp1}) using reductive perturbation technique \citep{K88}
\bq
\dtau{\bp} + \alpha\, \dxi{}\left[\left(|\bp|^2 - |b_{\perp 0}|^2\right)\,\bp\right]
+ \frac{i\,\eh}{2}\,\,\frac{\partial^2 \bp}{\partial \xi^2} = 0\,,
\label{eq:DN}
\eq
where $\bp = B/\Bz$ and $\alpha = v_A\,C_I/4\,(C_I^2 – C_s^2)$. One can show that Eq.~(\ref{eq:DN}) admits magnetically compressive (bright) and magnetically depressive (dark) solitons \citep{H98}.
Defining $S_{1,2} = \left(b_0 \mp \sqrt{2\,\epsilon/\alpha}\right)^2$ with
$\epsilon = \left(V/v_A –- C_I\right)/C_I$, describing the shift of the travelling wave speed $V$ from the intermediate speed and $\eta = \xi –- \epsilon\,\tau$ where $\xi = x –- C_I\,t$
and $\tau = C_I\,t$, the solutions can be written as
\begin{eqnarray}
S = S_0 + \frac{S_2 –- S_0}{1 + T_B\,\sinh^2\left(\alpha\,K\,\eta\right)}\,,\nonumber\\
\Phi = 3 \mbox{arctan}\left(C_B\,\tanh\left(\alpha\,K\,\eta\right)\right) + \nonumber\\
sgn \left( \frac{S_0}{2} - \frac{\epsilon}{\alpha}\right)
\mbox{arctan}\left(C_B^{*}\,\tanh\left(\alpha\,K\,\eta\right)\right), 
\nonumber\\
T_B = \frac{S_2 –- S1}{ S_0 –- S_1}\,,\,C_B = \sqrt{\frac{S_2 –- S0}{ S_0 –- S_1}}\,,
\end{eqnarray}
and $C_B^* = \sqrt{S_1/S_2} C_B$. Here $0<S_1 < S_0 < S_2$.
 \begin{eqnarray}
S = S_0 - \frac{S_0 –- S_1}{1 + T_D\,\sinh^2\left(\alpha\,K\,\eta\right)}\,,\nonumber\\
\Phi = - 3 \mbox{arctan}\left(C_D\,\tanh\left(\alpha\,K\,\eta\right)\right)  \nonumber\\
- sgn \left( \frac{S_0}{2} - \frac{\epsilon}{\alpha}\right)
\mbox{arctan}\left(C_D^{*}\,\tanh\left(\alpha\,K\,\eta\right)\right),
\nonumber\\
T_D = \frac{S_2 –- S1}{ S_2 –- S_0}\,,\,C_B = \sqrt{\frac{S_0 –- S1}{ S_2 –- S0}}\,,
\end{eqnarray}
and $C_D^* = \sqrt{S_2/S_1} C_D$ with $K = 0.5 \sqrt{\left(S_2 –- S_0 \right)\left(S_0 –- S_1\right)}$.
The value $b_1 = \sqrt{S_1}$ represents the minimum of the transverse field magnitude in a refractive, dark soliton
and $b_2 = \sqrt{S_2}$ is the peak value of $\bp$ in a compressive, bright soliton. The magnetic field depression in dark
maximizes when $b_1 = 0$, i.e. when transverse components $b_y$ and $b_z$ vanish. The magnitude of the field inside the
dark soliton is determined by $B_z = B_0\,\cos\theta$,  implying that a large propagation angle will cause large magnetic cavities.
The electric field for the magnetic cavities can be estimated as
\bq
E \sim B_0\,v_A\,\left(b_0 \mp \sqrt{\frac{2\,\epsilon}{\alpha}}\right)/c  \sin\theta\,.
\label{eq:EF}
\eq
Here $c$ is speed of light. Such a field is available for the charged particle acceleration. We speculate that if such a soliton induced field indeed exists in the solar photosphere, then with increasing height, since \alf speed will increase from $5.4\times 10^3\,\mbox{cm}\mbox{s}^{-1}$ at $h =0$ in the photosphere to $4\times 10^5\,\mbox{km}\mbox{s}^{-1}$ at $ h = 10^3\,\mbox{km}$ for a $10\,\mbox{G}$ magnetic field, the electric field $E$ will vary typically between $0.15 \,\mbox{V/m}$ to $1.2\,\mbox{V/m}$. Such a field may not be sufficient for accelerating the particles. However, if the field strength is increased by two to three orders of magnitude (as could be the case in the sunspots and pores) a million Volt field can easily exist that will be available for the acceleration and subsequent heating. We do not know of any observational support for the existence of soliton in the photosphere at this stage and thus the hypothesis of soliton generated field and ensuing charged particle acceleration should be treated with caution.

The following are the itemized summary of the present work.

(1) Non-ideal MHD effect, namely Hall diffusion is important in the solar atmosphere.

(2) The circularly polarized whistlers can be easily excited in the medium.

(3) The photosphere can become parametrically unstable.

(4) The nonlinear derivative Schr$\ddot{\mbox{o}}$dinger equation describes the finite amplitude fluctuations. We speculate that the dark and bright soliton solutions may cause the particle acceleration in the medium. 

\section*{Acknowledgments}

{BP wishes to thank Mark Wardle for his constant encouragement and support. The financial support of Australian Research Council and Macquarie University grant is acknowledged. It is our pleasure to thank the organisers of Kodai-Trieste Plasma Astrophysics workshop (August, 27 – September, 07), Kodaikanal, India which provided the stimulating atmosphere for developing part of this work.}

\appendix

\section[]{The coefficients of the dispersion relation Eq.~(18)}
Defining $b = \omega_0/ \omega_H$, $ a = B/B_z$ and $x = k /k_0$, the coefficients
of dispersion relation are given as
\begin{eqnarray}
a_8 = 1\,,\quad a_7 = 3\,b\,F_{\pm}\,x\,,\nonumber\\
%\end{equation}
%\begin{eqnarray}
\frac{a_6}{F_{\pm}} = b\,(1+x^2)\,\left[- 1 + b\,F_{\pm}\,(1-x^2)\right] - 2\,(1+x^2) - \frac{2}{F_{\pm}}
\nonumber\\
- x^2\,\left(\beta + a^2 - 2\,b^2\,F_{\pm} \right) \,,
\nonumber\\
\frac{a_5}{x\,F_{\pm}} = 2\,b\,F_{\pm}\,(1+ x^2) + \left(a^2 + 2\right)\left[1 - b\,F_{\pm}\,\left(1 - x^2
\right)\right]
\nonumber\\
+  1 - a^2 - 3\,b - b\,F_{\pm}\,x^2\,(\beta +  a^2)
%\nonumber\\
-  b\,\left[ 1 + F_{\pm}\,\left(1 + x^2\right)\right]
\nonumber\\
+ \frac{\left( 1 + x \right)}{x} - 2\,b\,\left[F_{\pm}\,(1 + x^2) + 1 + F_{\pm}\,x^2\,\left(\beta + a^2\right) \right]\,.
\end{eqnarray}
\begin{eqnarray}
\frac{a_4}{F_{\pm}} = \left(\beta + a^2\right)\,x^2 + F_{\pm}\,x^2\left[\beta\,\left(1+ x^2\right) + a^2\right]
+ a^2\,x^2 \nonumber\\
+ 2\,b\,x\left[F_{\pm}\,x\,\left(1 - a^2\right) - b\,x\,F_{\pm} - b\,x^3\,F_{\pm}^2\left(\beta + a^2\right) \right]
\nonumber \\
+ b\,x\,F_{\pm}\,\left(1 + x - 2\,b\,x\right) - 2\,x^2\,F_{\pm}\,\left(a^2 + 2\right)\nonumber\\
+ b\,\left(1+x^2\right)\left[ 1 - b\,F_{\pm}\,\left( 1 - x^2 \right) - F_{\pm}\,\left( 1 + x^2 \right)\right]
\nonumber\\
+ \left[ 1 + x^2 +  F_{\pm}^{-1} + x^2\,\left( \beta + a^2\right)\right]\left[ 1 + F_{\pm}\,\left( 1 + x^2\right)\right]
\nonumber\\
- \left[ 1 + x^2\left( 1 - a^2\right) - b \left(1 + x^2 \right) + F_{\pm}^{-1} -
b\,\beta\,F_{\pm}\,x^2 \right]\nonumber\\
\times \left(1 + x^2\right)\,\left[1 - b\,F_{\pm}\left( 1 - x^2\right)\right]\,.
\end{eqnarray}
\begin{eqnarray}
\frac{a_3}{F_{\pm}^2} = x\,\left( a^2 + 2 \right)\,\left[ -F_{\pm}^{-1} + b\,\left(1 - x^2\right)
+ \left(1 + x^2\right)  \right]+ 2\,x
\nonumber \\
\left[ 1 + x^2\,\left(1 - a^2\right) - b\left(1 +x^2 \right)\left(1 + \beta\,F_{\pm}\,x^2\right)
+ F_{\pm}^{-1} \right]\nonumber\\
- \left[F_{\pm}^{-1} - b\,\left(1 - x^2\right)\right]\left[F_{\pm}\,x^3\,\left(2\,\beta + a^2)\right)
+ 2\,x\,a^2 \right]
\nonumber\\
+ x^3\,\left(\beta + a^2\right)\left(b - 1\right) + 2\,b\,x\left[x^2\left(\beta + a^2\right) +
\beta\,F_{\pm}\,x^2
\right. \nonumber\\ \left.
\times \left(1 + x^2\right) + a^2\,x^2 \left(F_{\pm} + 1\right)\right] -
x\,\left[F_{\pm}^{-1} + 1 + x^2\right]\nonumber\\
\times \left[1 - b - a^2 - b\,F_{\pm}\,x^2\,\left( \beta + a^2\right) \right]
+ \left[ 2\,b\,x - 1 - x \right]
\nonumber\\
\times\left[1+ x^2 + F_{\pm}^{-1} + x^2\,\left(\beta + a^2\right)\right]\,.
\end{eqnarray}
\begin{eqnarray}
\frac{a_2}{F_{\pm}^2} = \left[ F_{\pm}^{-1} - b\,\left(1 - x^2\right) - 1 - x^2 \right]
\nonumber\\
\left[ 1 + x^2\, \left( 1 - a^2 \right)- b\, \left( 1 + x^2 \right)
\, \left( 1 + \beta\,F_{\pm}\,x^2 \right) + F_{\pm}^{-1} \right]
\nonumber\\
+  2\,x^2\,\left[F_{\pm}\,\x^2\,\left(2\,\beta + a^2\right) + 2\,a^2\right]
+ \left[ 1 - b\,F_{\pm}\, \left(1 - x^2\right)\right]
\nonumber\\
\times \left[ F_{\pm}^{-2} - b\,\beta\,x^2\,\left(1 + x^2\right) + x^2\,\left(\beta\,\left(1 + x^2\right)
- a^2\right)\right]
\nonumber\\
+ 2\,b\,F_{\pm}\,x^4\,\left( b -1 \right)\,\left( \beta + a^2\right) - x^2\,\left[F_{\pm}^{-1} + 1 + x^2 \right]
\nonumber\\
 \times \left[ \beta + a^2 + F_{\pm}\,\left(\beta\,\left(1 + x^2\right) + a^2\right) + a^2\right]
\nonumber\\
- x\,\left[2\,b\,x - 1 - x\right]\left[ 1 - a^2 - b - b\,F_{\pm}\,x^2\,\left(\beta + a^2\right)\right]\,.
\end{eqnarray}
\begin{eqnarray}
\frac{a_1}{F_{\pm}^2} = \left[ F_{\pm}^{-1} - b\,\left(1 - x^2\right) - 1 - x^2 \right]\times
\nonumber\\
\left[ x^2\,\left(2\,\beta + a^2\right)+2\, a^2\right] - 2\,F_{\pm}\,x\,
\left[ F_{\pm}^{-2} - b\,\beta\,x^2\,\left(1 + x^2\right)
\right. \nonumber\\ \left.
+ x^2\,\left(\beta\,\left(1 + x^2\right)
+ a^2\right)\right] - x^3\,\left(b - F_{\pm}\,x\right)\left(\beta + a^2\right)
\nonumber\\
\times \left[1 + F_{\pm}\left(1 + x^2\right)\right] + x^2\,\left( 1 + x - 2\,b\,x\right)\times
\nonumber\\
\left[\beta + a^2 + F_{\pm}\,\left(\beta\,\left(1 + x^2\right) + a^2\right) + a^2\right]\,.
\end{eqnarray}
\begin{eqnarray}
\frac{a_0}{F_{\pm}^2} = F_{\pm}\,x^3\,\left(b - 1\right)\,\left(\beta + a^2\right)\left( 1 + x - 2\,b\,x\right)
\nonumber\\
- \left[ F_{\pm}^{-2} - b\,\beta\,x^2\,\left(1 + x^2\right) + x^2\,\left(\beta\,\left(1 + x^2\right)
+ a^2\right) \right]
\nonumber\\
\times \left[ 1 - b\,F_{\pm}\,\left(1 - x^2\right) - F_{\pm}\,\left(1 + x^2\right)\right]\,.
\end{eqnarray}

\end{document}